\documentstyle[11pt,newpasp,twoside,epsf]{article}
\def\edcomment#1{\iffalse\marginpar{\raggedright\sl#1\/}\else\relax\fi}
\reversemarginpar
\begin{document}
\title{Type 1 AGN Unification}
\author{Martin Elvis}
\affil{Harvard-Smithsonian Center for Astrophysics,
60 Garden St.,  Cambridge MA02138 USA}

\begin{abstract}
%%%%%%%%%%%%%%%%
The model I recently proposed for the structure of quasars offers
to unify the many aspects of Type~1 AGN: emission lines,
absorption lines and reflection features.  This makes the model
heavily overconstrained by observation and readily tested.

Here I first outline the model, and then concentrate on how the
model answers objections raised since publication - with many of
the tests being reported at this meeting. I then begin to explore
how these and future tests can discriminate between this wind
model and 3 well-defined alternatives.

%%%%%%%%%%%%%%%%
\end{abstract}

%%%%%%%%%%%%%%%%%%%%%%%%%%%%%%%%%%%%%%%%%%%%%%%%%%%%%%%%%%%%%%%%%
\section{Introduction}

Just over 15 years ago AGN research made a major advance when
type~2 (narrow line) AGN were clearly identified as type~1 (broad
line) AGN viewed along a special, obscured, direction. This
unification of type~2 AGN with their unobscured cousins brought a
new degree of order and physical interpretation to the, then
recklessly multiplying, `types' of AGN, and a sense of the
importance of non-spherical geometry in these distant unresolved
objects.

Quasar research is still cursed with an overabundance of
phenomenology, even after `type~2 unification'. The strong
continuum of type~AGN has imprinted on it many different features
that encode aspects of the intimate environment of the nucleus:
narrow, broad and `very broad' emission lines, narrow
intermediate (`mini-BAL') and broad absorption lines, X-ray warm
and cold absorbers, optical/UV scatterers, X-ray reflection and
fluorescence features.  Unification of these details at all
wavelengths into a coherent picture has been little attempted,
and physics based explanations that include more than one
component are rare. As a result of the baffling richness of their
phenomenology quasars have become an unstimulating field for most
astronomers.

However stars were in a similar situation for at least 20 years
(c.1890-1911, Lawrence 1987). In fact the spectroscopic
definitions of the stellar types (O~B~A~F~G~K~M) read quite as
confusingly to outsiders as those of AGN classifications (e.g. G
stars: ``CaII strong; Fe and other metals strong; H weaker'',
Allen 1975).  We now know that the main sequence is a simple
temperature progression, determined fundamentally by stellar
mass. There is hope that the complexities of quasars will resolve
themselves the same way.  Using a 2-phase wind with a particular
geometry, I have proposed (Elvis 2000) a geometrical and
kinematic model for quasars that appears to subsume a great deal
of the phenomenology of quasar emission and absorption lines into
a single simple scheme, that is not without physical appeal.

Type~2 unification demonstrated that Quasars, unlike stars, are
not spherical. (In fact we have known that axisymmetry is
appropriate since the first double radio sources were discovered
(Jennison \& Das~Gupta 1953). This means that geometry matters,
and when this is the case the physics cannot be worked out until
we get the structure right: the solar system simply could not be
solved in a Ptolemaic geometry. A normal sequence in constructing
a physical theory is to work out the right geometry, then the
kinematics and lastly the dynamics (c.f. Copernicus - Kepler -
Newton). In quasars instead, I believe that the physics has
largely already been worked out, but discarded because the
geometry was not in place, making the physics appear
wrong. Axisymmetry is a crucial part of type~1 unification.

Here I briefly outline the model, and then concentrate on the
main objections that have been raised and respond to them.  Since
a model needs tests, and tests have to point to an alternative to
be strong, I have begun to explore alternative wind geometries to
see how their predictions differ from my model.

%%%%%%%%%%%%%%%%%%%%%%%%%%%%%%%%%%%%%%%%%%%%%%%%%%%%%%%%%%%%%%%%%
\section{A Structure for Quasars}

Winds are increasingly recognized as a common, perhaps
ubiquitous, feature of quasars and AGNs. Outflows at
$\sim$1000~km~s$^{-1}$ are directly seen in absorption in half of
all AGN (Hutchings et al. 2001, Kriss 2001, Crenshaw \& Kraemer
1999, Reynolds 1997), while the much faster,
$\sim$5000-10,000~km~s$^{-1}$, winds of Broad Absorption Line
(BAL) quasars must be present in much more than the observed 10\%
of quasars, since the strongly polarized emission in the BAL
troughs (Ogle 1997) requires a highly non-spherical structure.

In Elvis (2000) I proposed that a flow of warm ($\sim 10^6$K) gas
rises vertically from a {\em narrow range of radii} on an
accretion disk. This flow then accelerates, angling outward (most
likely under the influence of radiation pressure from the intense
quasar continuum) until it forms thin conical wind moving
radially (figure~1).  When the continuum source is viewed through
this wind it shows narrow absorption lines (NALs) in both UV and
X-ray (the X-ray `warm absorber'); when viewed down the radial
flow the absorption is stronger and is seen over a large range of
velocities down to $v(vertical)$, the `detachment velocity', so
forming the Broad Absorption Line (BAL) quasars. Given the
narrowness of the vertical flow ($\sim 0.1r$), the divergence of
the continuum radiation at the turning point will be
$\sim$6$^{\circ}$, giving 10\% solid angle coverage, and so the
correct fraction of BAL quasars.  [The angle to the disk axis,
60$^{\circ}$, is at present arbitrarily chosen to give the
correct number of NAL and non-NAL quasars.]

%%%%%%%%
\begin{figure}
%\plotfiddle{fig1_bw.eps}{3in}{0}{45}{45}{-150}{0}
\plotfiddle{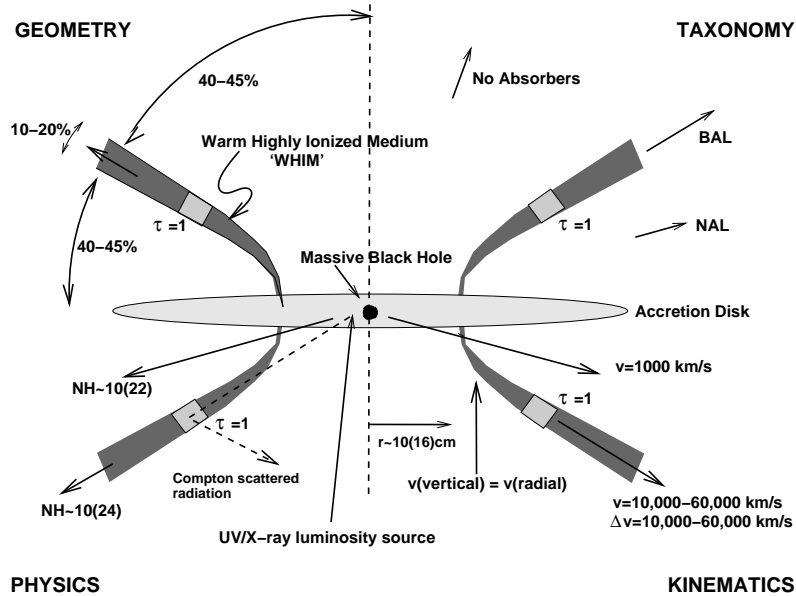}{3in}{0}{45}{45}{-150}{0}
\caption{A Structure for Quasars (Elvis 2000).}
\end{figure}
%%%%%%%%

This `Warm Highly Ionized Gas' (WHIG
\footnote {Initially I used `WHIM' (i.e. `medium' instead of
`gas') but this acronym has also been used for the `Warm-Hot
Intergalactic Medium' (Cen \& Ostriker 1999), forcing me to
adopt WHIG instead.}
) has a cool phase (like the ISM) with which it is in pressure
equlibrium. This cool phase provides the clouds that emit the
Broad Emission Lines (BELs). Since the BEL clouds move along with
the WHIG they are not ripped apart by shear forces; and since the
medium is only Compton thick along the radial flow direction
rapid continuum variations are not smeared out. Both problems had
long been a strong objections to pressure confined BEL clouds,
but they are invalid in the proposed geometry.

The radial flow is Compton thick along the flow direction, and so
will scatter all wavelengths passing along that direction. Since
the flow is highly non-spherical the scattered radiation will be
polarized. The solid angle covered by the radial flow is
10\%-20\%, so this fraction of all the continuum radiation will
be scattered, leading to the filling in of the BAL troughs, and
to an X-ray Compton hump in all AGN. Since the WHIG is only
ionized to FeXVII, there will be Fe-K fluorescence off the same
structure at $\sim$100~eV EW. Some of the BEL radiation will also
pass along the flow and will be scattered off the fast moving
flow, producing the polarized, non-variable `Very Broad Line
Region'.

%%%%%%%%%%%%%%%%%%%%%%%%%%%%%%%%%%%%%%%%%%%%%%%%%%%%%%%%%%%%%%%%%%
\section{Key Criticisms \& Responses}

Several criticisms of the Elvis (2000) model have come up more
than once and bear careful consideration. 

\medskip
%%%%%%%%%%%%%%%
\noindent$\bullet${\em Where are the BAL Seyferts?}

BALs have not been found in HST studies of low luminosity AGN,
but only in quasars. This would suggest a luminosity dependent
wind velocity. How can this observation fit with the model?  The
simplest possibility is that {\em BALs are present at low~L but
had been missed}. To qualify as a BAL an absorption line needs to
have a `velocity spread' $>$2000~km~s$^{-1}$ (Weymann et
al. 1991).  In fact, of 34 type~1 AGN observed with FUSE (Kriss
2001) two have an OVI absorber with FWHM$>$2000~km~s$^{-1}$.
This fraction is in reasonable agreement with the expected
$\sim$10\% fraction, given the limited statistics.

Two other explanations require that low luminosity AGN are subtly
different from high luminosity AGN. e.g. their BALs may be too
highly ionized. Since BALs are now known to have high ionization
(Telfer et al. 1998, Ogle 1998), only a small increase in $U$
(from $U\sim 2$ to $U\sim 6$) would remove CIV altogether (i.e. a
factor $>$100 reduction in CIV ion fraction, Mathur et al., 1995,
fig.~4).  The original suggestion of Elvis (2000) was that BAL
outflows become dusty at low luminosities.

The first option is a clear favorite, as it is predicted by the
model and so requires no additional free parameters.  To be sure
a much larger sample is needed. Fortunately a sample of 80 low
luminosity AGN is planned to be observed with FUSE (Kriss 2001).

\medskip
%%%%%%%%%%%%%%%
\noindent$\bullet$ {\em UV absorbers don't have sufficient column
density to be the X-ray warm absorbers.}

Some analyses (Kriss et al. 1966) do find this to be the case,
while others do not (Shields \& Hamann 1997). However, Kaspi et
al. (2001) note that this analysis is critically dependent on the
(unobserved) EUV continuum. Kaspi et al. can match UV and X-ray
column for NGC~3783 for a particular, well-constrained continuum
form is assumed. This sensitivity is a strength, since accretion
disk models predict specific EUV continuum shapes, which will be
stringently tested if the X-ray/UV absorbers are the
same. Moreover NLR optical coronal lines also have to have ratios
that agree with this continuum shape.

More important is the observation by Arav, Korista \& de~Kool
(2001) that, although the UV NALs {\em appear} to be unsaturated,
the conventional curve-of-growth based analysis (Mathur et
al. 1995 [MEW95], Kaspi et al. 2001) cannot apply to OVI or
Ly$\alpha$.  In several AGN the OVI doublet ratio and
Ly$\alpha$/Ly$\beta$ ratio do not match the ratio of their
respective oscillator strengths, as expected in the optically
thin case. Instead Arav et al. show that the residual flux in the
absorption line troughs must either be unobscured continuum
produced through scattering back into our line of sight, or
simply due to the absorbing gas not covering the whole continuum
source.  Overlooking this possibility is the same understandable
mistake that led to BAL column densities underestimated by
factors of 100-1000 for many years (Mathur, Elvis \& Singh 1995,
Ogle et al., 1998). In NGC~5548 Arav et al.  show that the true
column density of CIV is at least four times larger than a simple
curve-of-growth analysis would give. Consistency with the X-ray
columns is then readily achievable.

The result is a far simpler picture of AGN with only one, high
ionization, absorber, and a realization that the FUSE spectra
give us far more information about the geometry of quasar winds,
and potentially on the shape of the continuum source, that we
could have hoped.

\medskip
%%%%%%%%%%%%%%%
\noindent$\bullet${\em The X-ray emission lines are narrower than
the optical/UV BELs:}

The emission lines seen around 1~keV in {\em Chandra} and
XMM-Newton AGN spectra have widths of $\leq$1000~km~s$^{-1}$,
several times narrower than the BEL widths(Krolik \& Kriss 2001).
They are consistent with an origin in the Narrow Line Region
(Ogle et al. 2000), or the hypothesized `donut'-shaped torus.

However BEL-shaped X-ray lines would not be detectable in the
existing {\em Chandra} or XMM-Newton spectra because BEL widths
of $\sim$3000~km~s$^{-1}$ are over-resolved in {\em
Chandra}. Figure 2 shows a simulation of a long {\em Chandra}
HETG spectrum with narrow X-ray emission lines from similar to
those in Mrk~3 and NGC~4151 (Sako et al., 2000, Ogle et
al. 2000), and broad lines matching the NGC~5548 BEL widths for a
covering factor 0.1 (F. Nicastro 2001, private communication).
The broad lines are not detectable.
%%%%%%%%
\begin{figure}
\plotfiddle{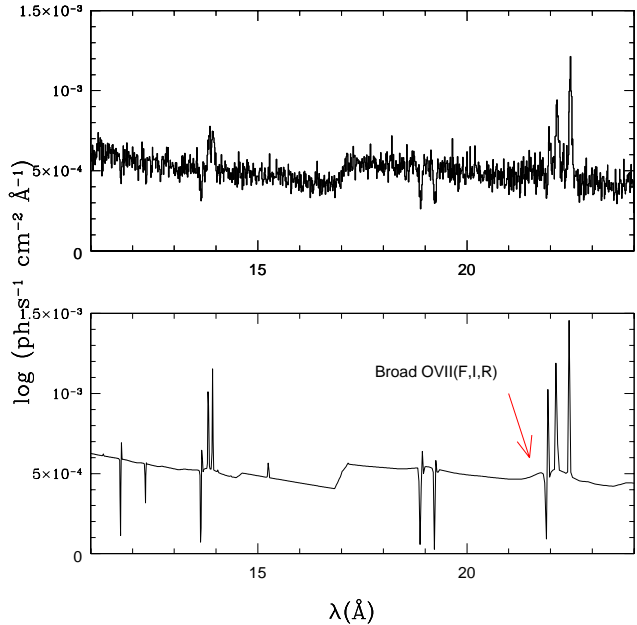}{3in}{0}{45}{45}{-150}{-80}
%\plotfiddle{sim.ps}{3in}{0}{45}{45}{-150}{-80}
%                   h       %x  %y  dx   dy
%\plotone{sim.ps}
\caption{(a) Simulated 750~ksec {\em Chandra} HRC-LETGS spectrum
of NGC~5548 including a BLR component ($f_C$=0.1; (b) input
spectrum, the broad OVII line is marked). F.Nicastro (2001),
private communication.}
\end{figure}
%%%%%%%%

Much longer exposures with {\em Chandra} could provide strong
limits. The expected WHIG line profiles then need to be
calculated carefully, since the WHIG is optically thin to the
X-ray lines, but BEL clouds are optically thick. This means that
the profiles of X-ray and optical broad lines will be different,
with the X-rays probably showing a double peaked structure. This
line shape also may confuse present analysis.

\medskip
%%%%%%%%%%%%%%%
\noindent$\bullet$ {\em UV absorbers have multiple velocity
components requiring multiple absorbing zones, probably with
different ionization states.}

X-ray instrumentation is sensitive to only a limited range of
column densities, while ultraviolet spectra can only detect a
limited range of high ionization states.  For comparable
ionization states systems with column densities at least 10 times
smaller than the X-ray (Kriss et al. 2000) can be detected by HST
and FUSE spectra than by {\em Chandra}
(N$_{H~min}~\sim$10$^{20.5}$cm$^{-2}$, Collinge et al. 2001),
since {\em Chandra} is limited in both S/N and resolution
(Collinge et al. 2001).  High ionization aborption systems
instead can only be seen in X-ray spectra, since the ion
populations of the UV transitions become tiny. (e.g. the high
velocity system in NGC~4051, Collinge et al. 2001.) Modest
inhomogeneities in the WHIG wind could produce the needed
differences in $U$.

It is worth recalling that the `single X-ray/UV absorber' model
has already passed two tests. Before the {\em Chandra} high
resolution grating spectra it was possible that the X-ray
absorbers had quite different velocities and widths to the UV
absorbers.  Instead the {\em Chandra} spectra confirmed a good
correspondence between the two systems in both redshift and line
width (Kaspi et al. 2001, Collinge et al. 2001), which was a
strong prediction of the model. Radiatively accelerated winds
seem to be inherently unstable and normally produce highly
structured line profiles (e.g. P~Cygni, Stahl et al. 1993). The
detailed sub-structure is then `mere' weather.

\medskip
%%%%%%%%%%%%%
\noindent$\bullet${\em BALs are too low ionization to be due to
the X-ray Warm Absorber gas.}

Indirect arguments based on the large X-ray column densities of
BALs (MES95) argue for high ionization. More recently BALs have
been found in the phosphorus PV ion, confirming unequivocally
their high ionization, although so far only for a few quasars
(Hamann et al., 1995).

Since Compton scattering is wavelength independent the BAL
covering factor in the soft X-ray band should be the same as in
the optical if the BAL material is sufficiently ionized.  Partial
covering of the right order is seen in the X-ray spectrum of one
BAL quasar (Mathur et al. 2001).  The low energy X-rays should
also be polarized like the optical BAL troughs, an observation
that may now become feasible (daCosta et al. 2001).

\medskip
%%%%%%%%%%%%%
\noindent$\bullet${\em The X-ray absorbers are too far from the
BELR}

The radius of the NAL from the continuum source has to match the
radius of the high ionization BELR. In NGC~5548 this works, but
the NAL radius is poorly constrained (a factor 100). Short
recombination times, which will determine the NAL density, are
the key to a better constraint on the radius. Timescales of hours
are likely, and require quite large continuum variations on this
timescale or less to be measurable.

There are also high ionization BELs of PVII and NeVIII now
detected at $\sim$770\AA (rest) in some high redshift quasars
(Hamann et al. 1995). These do not match normal BEL cloud
conditions, but could arise from the WHIG. In this case they
should match the WHIG properties.  In fact Hamann et al. (1995)
find values that do fit: column densities of
$\sim$10$^{22}$cm$^{-2}$, a temperature of 5$\times$10$^{5}$K (if
collisional), and a covering factor of 0.5 like that of NALs (and
unlike the BELR, for which 0.1 is a typical value).

\medskip
%%%%%%%%%%%%%%%
\noindent$\bullet${\em Why should the wind be narrow? What
`special' radius could there be?}

We know that there is some physics of accretion disks that causes
instabilities over a relatively narrow range of radii since QPOs
in X-ray binaries have a width $\Delta\nu/\nu\sim$0.1. If
these QPOs arise in accretion disks then they have a spread in
radius of the same order. So we have good evidence that special
ranges of radii can exist in accretion disks, even though the
physical cause is still debated, even in the case of QPOs
(B. Czerny 2001, private communication).

The NALs are stable features of AGN spectra over at least 20
years (MEW95). This might have been because they lie a long way
from the continuum source, but (in NGC~5548, Shull and Sachs
1993) the short CIV recombination time ($<$4~days) requires a
high density ($n_e>5\times$10$^4$cm$^{-3}$). Given the high
ionization parameter ($U\sim$1) the absorber cannot be more than
2$\times$10$^{18}$cm from the continuum source (MEW95) and no
more than $\sim$10$^{16}$cm thick. The crossing time for a
spherical blob of this radius is 10$^8~M_6^{-\frac{1}{2}}$~s at
10$^{18}$cm, requiring a cloud elongated by at least 6:1 to be
seen for 20 years, implying a length nearly 0.1 the distance from
the center.
A higher density cloud will be smaller and nearer the center (to
maintain $U$) and will have shorter crossing times
($\tau_{c.cross}\propto n_e^{-\frac{5}{4}}$). The concept of
obscuring clouds then morphs into that of a continuous flow with
an elongation equal to the radial distance at a density $\sim$40
times higher. An improved density estimate for the absorber (and
hence the distance from the center), from UV or X-ray monitoring,
will give a clear test.

The timescales for stochastic flares (as seen in initial
simulations, e.g. Proga 2001) need to be investigated to see if
they can produce stable absorbers of sufficient density over
decade long timescales.  If not then some more continuous
injection/ejection process will be needed.

\medskip
%%%%%%%%%%%%%%%
\noindent$\bullet${\em The BELR is not in radial motion.}

Reverberation mapping (Peterson 1997) shows that the BELR is not
dominated by radial motion, either outflow or inflow. However,
this is not the prediction of the model. The initial flow from
the accretion disk is vertical, producing a thin cylinder that
has a primarily rotational velocity at the disk Keplerian speed
(see also Nicastro 1999). Laor (2001) finds that weak EW([OIII]),
the presence of NALs, and width of CIV are correlated in PG
quasars. If [OIII] is isotropic and the continuum is from a disk
then the EW([OIII]) is an inclination indicator, and broader CIV
with absorbers will be found in edge-on AGN if the BELR is
disk-like or cylindrical (Risaliti, Laor \& Elvis 2001, in
preparation).  N. Murray (2001, private communication) notes that
a cylinder or disk geometry for the BELR is suggested by the
observation of double-peaked profiles for CIV following a flare
in NGC~5548, when the BELs may become optically thin.

A cylindrical BELR may also explain the peculiar rotation of
polarization position angle of H$\alpha$ in AGN (Cohen \& Martel
2001, Axon, these proceedings).  BEL photons will scatter off the
opposite side of the rotating cylinder. A cylinder, being
hollow, is optically thin to BELR photons, allowing them to cross
readily to the other side.  Similar PA rotations are implied for
{\em all} edge-on AGN , i.e. those with NALs. Axon (these
proceedings) shows that AGN have two types of H$\alpha$
polarization, apparently consistent with the two main viewing
angles of a cylindrical BELR. His data promise strong tests of
the geometry.

\medskip
%%%%%%%%%%%%%%%
\noindent$\bullet${\em The narrow Fe-K line comes from the torus,
not a BAL wind.}

A `narrow' Fe-K line is seen in most Seyfert galaxies, and this
is usually ascribed an origin in the pc-scale `donut'-shaped
torus normally employed in type2/type~1 unification models (Urry
\& Padovani 1995). This Fe-K line provides a strong test of the
model since the BAL-producing radial outflow {\em must} produce
Fe-K emission. This outer, conical, part of the WHIG has to be
Compton thick to produce the scattered light seen in BAL troughs
(Ogle 1998). Since Compton scattering is grey (wavelength
independent) the same region will produce a symmetric X-ray
6.4~keV Fe-K emission line.  The Fe-K line from the structure
will have a width comparable to, but broader than, the BELR
widths (since the wind is accelerating outward). The line should
be a factor $\sim$2 broader if the wind is radiatively
accelerated. So the Fe-K width is likely to be correlated with
the BEL widths, and will be clearly broader than the NLR-like
widths associated with a `donut' shaped torus, but clearly
narrower than the relativistic widths that would be produced by
the inner regions of an accretion disk.

Other parts of an AGN (a `donut', the NELR or an accretion disk)
may also produce Fe-K. High resolution X-ray spectra (e.g. with
the HETGS on {\em Chandra}) will be able to resolve these
components, and so determine whether a WHIG-related Fe-K line
exists as predicted.  The first Chandra HETG results seem to
indicate that such fairly broad (BELR-like) components of Fe-K do
exist: NGC~5548 has a resolved Fe-K line
FWHM=4525$^{+3525}_{-2645}$~km~s$^{-1}$(90\%, Yaqoob et
al. 2001 , c.f. H$\beta$ FWHM=3700~km~s$^{-1}$, Osterbrock 1977);
while NGC~3783 has FWHM$<$3250~km~s$^{-1}$ (Kaspi et al., 2001, ,
c.f. H$\beta$ FWHM=4100$\pm$1160~km~s$^{-1}$, Wandel et
al. (1999). 

Variability gives another clean cut way to discriminate between
Fe-K from a pc-scale `donut', BELR-related emission, and the
inner regions of an accretion disk. Several studies have already
shown that the `narrow' Fe-K component does not vary in unison
with the X-ray continuum (e.g. Chiang et al. 2000, Risaliti
2001), ruling out the accretion disk origin. A BELR-related Fe-K
line would follow the X-ray continuum with a reverberation delay
larger than but similar to the CIV reverberation delay.  Takashi,
Inoue \& Dotani (2001) have found such an effect in
NGC~4151. They derive an Fe-K scattering region size of
10$^{17}$cm which is five times larger than the CIV radius for
NGC~4151 (9$\pm$2~light-days, 2$\times$10$^{16}$cm, Kaspi et
al. 1996), consistent with the proposed structure. This size is
inconsistent with a pc-scale `donut' origin. This is strong
support for the proposed structure.

The optical/UV continuum is scattered by the structure.  At any
given angle there will be four dominant delay times relative to a
central continuum flaring event, as the flare scatters off the
near and far parts of the $\tau$=1 rings above and below the disk
plane. (The disk may obscure one or both parts of the lower
ring.) These should show up in autocorrelation functions of the
continuum at low amplitudes.  Schild (2001) has found that
previously published (Schild 1986) autocorrelation timescales in
the gravitational lens Q~0957+561 are consistent, in an
overconstrained set of equations, with the double ring expected
from this structure.

The X-ray `Compton Hump' should show the same time smearing and
delays as the Fe-K line. Moreover the Compton Hump should be
polarized, just as the optical BAL troughs are polarized. DaCosta
et al. (2001) have just demonstrated a high efficiency X-ray
polarimeter. Behind a reasonable sized X-ray mirror this could
make clear tests of the model.

%%%%%%%%%%%%%%%%%%%%%%%%%%%%%%%%%%%%%%%%%%%%%%%%%%%%%%%%%%%%
\section{Alternative Wind Models}

To reconcile the presence of both a fast (BAL) wind and a slow
(NAL) wind in most AGN and quasars, we adopted a single wind with
a special geometry. It is worth considering alternative means of
reconciling these two winds by relaxing the constraint that they
are the same flow. There are two main possibilities: either high
luminosity objects have faster winds, or faster winds are emitted
in some preferential direction, and slower winds in others. Here
we begin to examine these options. We assume that the models
retain all the other features of our model. I.e. they still try
to combine the UV and X-ray absorbers in a single WHIG (our
starting point); and they have the BEL clouds embedded in this
wind; they try to explain reflection features via scattering off
the fast wind.

%%%%%%%%%%%%%%%%%%%%
\subsection{Luminosity Dependent Velocities}

In this hypothesis the fast BAL winds are emitted only by high
luminosity quasars (into a $\sim$10\% solid angle), while NAL
winds are found only at lower luminosities (for a $\sim$50\%
solid angle). The wind no longer needs to arise from a narrow
range of disk radii. 

There are a number of comments that can be made:

\noindent (1) In this scenario a continuum of widths should be
found with a covering factor that decreases from $\sim$50\% at
NGC~5548 luminosities (L$_X\sim$5$\times$10$^{43}$erg~s$^{-1}$,
Elvis et al., 1978), to $\sim$10\% at
L$_X\sim$10$^{46}$erg~s$^{-1}$ PHL~5200, MES95).  There is a
range of BAL velocities, but a line width vs. luminosity or
covering factor analysis has not yet been performed.

\noindent (2)
This model has a built-in explanation for the absence of low
luminosity BALs.

\noindent (3) This model has no obvious physical cause for a
BAL `detachment velocity'.

\noindent (4) Can high ionization BELs with large covering factor
arise in this model?

\noindent (5) The scattering effects of the radial part of the
wind (X-ray Fe-K emission line, Compton hump, optical polarized
flux) will disappear in lower luminosity objects since the column
density needed to produce scattering is much larger than the
column density through the X-ray warm absorbers, and low
luminosity AGN will have no fast wind out of the line of sight to
produce the polarized, non-variable `Very Broad Line Region'.
The opposite is observed (Iwasawa \& Taniguchi 1993).

\noindent (6)
The BAL opening angle of 10\% does not arise naturally from the
narrow width of the wind origin site on the disk.

\noindent (7)
Are there directions in which one looks through the wind?
If so then NALs will be seen, the wind originates from a
restricted range of radii and the picture reverts to something
close to our model.

The absence of reflection effects in this version considerably
weakens its unifying power.

%%%%%%%%%%%%%%%%%%%%
\subsection{Orientation Dependent Velocities}

In this hypothesis the quasar wind has fast (BAL) velocities only
in some directions (covering $\sim$10\% solid angle), and has
slower (NAL) velocities in other directions (covering $\sim$50\%
solid angle). There are two obvious preferential directions for
the fast wind: {\em polar} and {\em equatorial} (figure 3).

%%%%%%%%
\begin{figure}
\plotfiddle{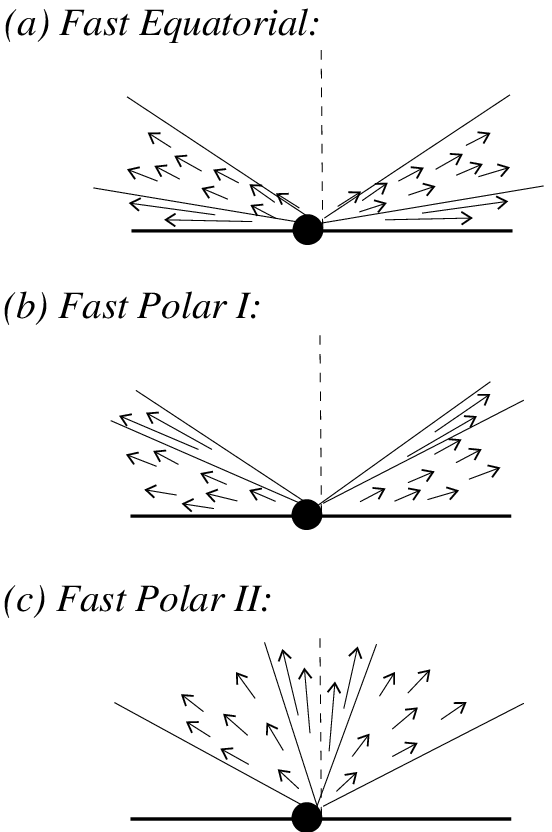}{3in}{0}{100}{100}{-70}{0}
%\plotfiddle{fig3.eps}{3in}{0}{100}{100}{-70}{0}
%\plotone{fig3.eps}
\caption{Options for quasar winds with direction dependent velocities.}
\end{figure}
%%%%%%%%

Some comments are:

\noindent (1)
Proga (these proceeedings) finds such a directional velocity
stratification arising naturally from his simulations, with
higher velocities toward the pole (figure 3b).

\noindent (2) 
Like our model, this hypothesis does not explain
the absence of low luminosity BALs, but they may in fact exist
(see \S3).

\noindent (3) 
Again, this model has  no obvious physical cause for a
`detachment velocity'. 

\noindent (4) Compton scattering features will arise naturally in
all objects in this model, as in ours, if the fast wind has
sufficient column density.

\noindent (5) The fast wind has to have a large column density of
high velocity material when viewed end-on to reproduce the BAL
observations. This arises naturally in our model, but is not
obvious in this hypothesis. A large column density implies a
large mass input rate at the wind base, decaying rapidly to
larger radii (in the fast polar case; to smaller radii for the
fast equatorial case).

\noindent (6) In the fast polar case a much ($<$1\%) lower column
density is required when viewed from other directions to avoid
BALs being seen more often. A long thin BAL region is necessary,
which would have to be non-divergent to maintain column density.
This may make it difficult to cover a 10\% solid angle. An
equatorial fast wind avoids this problem.

\noindent (7) 
An equatorial fast wind has the faster material originating
further from the continuum source, which seems unlikely since
radiation pressure and Keplerian rotation both predict the
opposite. A polar fast wind avoids this problem.

The difficulties in this version of the model lie
primarily in theory: can the large column densities
needed in the fast wind be produced, in an acceptable geometry?
Neither the polar nor equatorial solution is fully appealing.

%%%%%%%%%%%%%%%%%%%%%%%%%%%%%%%%%%%%%%%%%%%%%%%%%%%%%%%%%%%%
\section{Conclusions}

Because the Elvis (2000) model unifies so many aspects of Type~1
AGN phenomenology it is highly overconstrained, and so readily
tested. This is a strength of the model. The model has passed
quite a number of tests already, but they are not yet as
stringent as one would like.

Quite simple extensions of the model (e.g. to type~2 objects,
Risaliti, Elvis, \& Nicastro 2001) suggest that much more of the
quasar/AGN phenomenology can be incorporated with only a handful
of extra variables.  Consideration of the fate of the quasar
wind, which clearly exceeds the escape velocity of any galaxy or
cluster of galaxies, should also be illuminating (e.g. Elvis,
Marengo \& Karovska 2001). With luck then, quasars will now enter
a period of rapid development of their physics (e.g. Nicastro
2000), allowing their physical evolution to be understood, and
placing them constructively within cosmology.

\smallskip
I thank Nahum Arav, Bo\.{z}ena Czerny, Fred Hamann, Norm Murray
and Fabrizio Nicastro for enlightening discussions.
This work was supported in part by NASA grant NAG5-6078 (LTSA).

%%%%%%%%%%%%%%%%%%%%%%%%%%%%%%%%%%%%%%%%%%%%%%%%%%%%%%%%%%%%

%%%%%%%%%%%%%%
\end{document}